\documentclass{article}%
\usepackage{amsmath}
\usepackage{cite}
\usepackage{citesort}
\usepackage{graphicx}%
\usepackage{amsfonts}%
\usepackage{amssymb}

\begin{document}
\thispagestyle{empty}  \setcounter{page}{0}  \begin{flushright}%

May 2007 \\
\end{flushright}

\vskip                                         3.5 true cm

\begin{center}
{\huge Improved estimates of rare $K$ decay}\\[5pt]

{\huge matrix-elements from $K_{\ell3}$ decays}\\[28pt]

\textsc{Federico Mescia}$^{1}$ \textsc{and Christopher Smith}$^{2}$
\\[30pt]$^{1}~$\textsl{INFN, Laboratori Nazionali di Frascati, I-00044
Frascati, Italy}\\[5pt]$^{2}~$\textsl{Institut f\"{u}r Theoretische Physik,
Universit\"{a}t Bern, CH-3012 Bern, Switzerland }\\[55pt]

\textbf{Abstract}
\end{center}

\noindent The estimation of rare $K$ decay matrix-elements from $K_{\ell3}$
experimental data is extended beyond LO in Chiral Perturbation Theory.
Isospin-breaking effects at NLO (and partially NNLO) in the ChPT expansion, as
well as QED radiative corrections, are now accounted for. The analysis relies
mainly on the cleanness of two specific ratios of form-factors, for which the
theoretical control is excellent. As a result, the uncertainties on the
$K^{+}\rightarrow\pi^{+}\nu\bar{\nu}$ and $K_{L}\rightarrow\pi^{0}\nu\bar{\nu
}$ matrix-elements are reduced by a factor of about $7$ and $4$, respectively,
and similarly for the direct CP-violating contributions to $K_{L}%
\rightarrow\pi^{0}e^{+}e^{-}$ and $K_{L}\rightarrow\pi^{0}\mu^{+}\mu^{-}$.
They could be reduced even further with better experimental data for the
$K_{\ell3}$ slopes and the $K_{\ell3}^{+}$ branching ratios. As a result, the
non-parametric errors for $\mathcal{B}\left(  K\rightarrow\pi\nu\bar{\nu
}\right)  $ and for the direct CP-violating contributions to $\mathcal{B}%
\left(  K_{L}\rightarrow\pi^{0}\ell^{+}\ell^{-}\right)  $ are now completely
dominated by those on the short-distance physics.\newpage

\section{Introduction}

The rare decays $K\rightarrow\pi\nu\bar{\nu}$ and $K_{L}\rightarrow\pi^{0}%
\ell^{+}\ell^{-}$ are important tools to test the Standard Model (SM), and to
search for possible new physics. As they proceed through flavor changing
neutral currents (FCNC), though they are very suppressed in the SM, they show
an exceptional sensitivity to short-distance physics. Recent theoretical
progress has greatly improve the theoretical control over the SM
predictions\cite{BurasGHN05,IsidoriMS05,Kpill}. On the experimental side, rare
$K$ programs at CERN and J-PARC are currently under study, aiming at a
measurement by the beginning of the next decade.

An essential aspect of these decay modes is of being semi-leptonic. Contrary
to $\varepsilon^{\prime}/\varepsilon$ for example, the dominant contribution
does not come from four-quark operators, but rather from quark vector-current
FCNC operators like $(\bar{s}\gamma_{\mu}d)(\bar{\nu}\gamma^{\mu}\left(
1-\gamma_{5}\right)  \nu)$, on which we have an excellent control. This stems
from their relationship under the isospin symmetry with the charge-current
(CC) Fermi operators $(\bar{s}\gamma_{\mu}u)(\ell\gamma^{\mu}\left(
1-\gamma_{5}\right)  \bar{\nu}_{\ell})$. More specifically, the FCNC hadronic
matrix-elements required for rare $K$ decays as well as the one for the CC
transition in $K\rightarrow\pi\ell\nu_{\ell}$ decays ($K_{\ell3}$) are
parametrized as $\left(  T=P-K\right)  $%
\begin{equation}
\langle\pi^{j}\left(  K\right)  |\bar{q}\gamma^{\mu}\lambda_{a}q|K^{i}%
(P)\rangle=C_{ij}(f_{+}^{K^{i}\pi^{j}}(T^{2})\left(  P+K\right)  ^{\mu}%
+f_{-}^{K^{i}\pi^{j}}(T^{2})\left(  P-K\right)  ^{\mu})\;,
\end{equation}
with $C_{ij}$ some Clebsch-Gordan coefficients and the Gell-Mann matrices
$\lambda_{FCNC}=\lambda_{6}\pm i\lambda_{7}$, $\lambda_{CC}=\lambda_{4}\pm
i\lambda_{5}$ projecting out the desired quark-flavor structures. In the
isospin limit, all these form-factors are equal.

At lowest order in Chiral Perturbation Theory (ChPT), these form-factors are
all related to the conserved current form-factors $f_{\pm}^{\pi^{+}\pi^{-}%
}(T^{2})$, and thus $f_{+}^{K^{i}\pi^{j}}(T^{2})=1$, $f_{-}^{K^{i}\pi^{j}%
}(T^{2})=0$ for all $i,j=+,0$. Further, the Ademollo-Gatto theorem protects
against large $SU\left(  3\right)  $ corrections, which can arise only at the
second order in $m_{s}-m_{u,d}$. Of course, in practice, it is not useful for
us to compute these $SU\left(  3\right)  $ corrections since they are the same
for all $K\rightarrow\pi$ transitions. Indeed, our goal is to use instead the
precise experimental information on the form-factors obtained from $K_{\ell3}$
decays. The problem then reduces to the study of isospin-breaking effects,
proportional to $\varepsilon^{(2)}\sim(m_{u}-m_{d})/m_{s}$, to relate FCNC and
CC form-factors as precisely as possible.

This strategy of using $K_{\ell3}$ data is common practice, but currently
relies on the $\mathcal{O}(p^{2}\varepsilon^{(2)})$, LO analysis of
Ref.\cite{MarcianoP96}. Given the recent theoretical progress in the
computation of short-distance QCD corrections, it is now time to improve and
go beyond LO. More precisely, with respect to Ref.\cite{MarcianoP96}, our
goals are:

\begin{quote}
1 -- To include isospin-breaking effects at NLO (and partially NNLO) in the
ChPT expansion.

2 -- To account for QED radiative corrections, at leading order in the ChPT expansion.

3 -- To update matrix-elements using the latest $K_{\ell3}$ experimental data.

4 -- To perform a detailed error study, including both theoretical and
experimental uncertainties.
\end{quote}

The outline of the paper is as follows. In the next Section, the master
formulas are given. In Section 3, the ChPT results for the form-factors at
$\mathcal{O}(p^{4}\varepsilon^{(2)})$, and partially $\mathcal{O}%
(p^{6}\varepsilon^{(2)})$, are given and discussed. Special emphasis will be
set on two ratios of form-factors on which an exceptional theoretical control
can be reached. Then, the numerical analysis is performed in Section 4 and our
results are summarized in the Conclusion. Finally, loop functions as well as
the details of the computation of the QED radiative corrections are presented
in the Appendix.

\section{Generalities}

The FCNC weak Hamiltonian relevant for the $K\rightarrow\pi\nu\bar{\nu}$
decays and for the direct CP-violating contribution to the $K_{L}%
\rightarrow\pi^{0}\ell^{+}\ell^{-}$ decays is%
\begin{gather}
H_{eff}=\frac{G_{F}\alpha\left(  M_{Z}\right)  }{\sqrt{2}}\sum_{\ell
=e,\mu,\tau}\left(  \frac{y_{\nu}}{2\pi\sin^{2}\theta_{W}}Q_{\nu\bar{\nu}%
}\right)  -\frac{G_{F}\alpha\left(  M_{Z}\right)  }{\sqrt{2}}\sum_{\ell=e,\mu
}\left(  y_{7V}Q_{7V}+y_{7A}Q_{7A}\right)  +\mathrm{h.c.\;,}\\
Q_{\nu\bar{\nu}}=\left(  \bar{s}\gamma_{\mu}d\right)  \times\left(  \bar{\nu
}_{\ell}\gamma^{\mu}\left(  1-\gamma_{5}\right)  \nu_{\ell}\right)
,\;\;Q_{7V}=\left(  \bar{s}\gamma_{\mu}d\right)  \times\left(  \bar{\ell
}\gamma^{\mu}\ell\right)  ,\;\;Q_{7A}=\left(  \bar{s}\gamma_{\mu}d\right)
\times\left(  \bar{\ell}\gamma^{\mu}\gamma_{5}\ell\right)  \;.
\end{gather}
In the SM, the Wilson coefficient $y_{\nu}$ is given by%
\begin{equation}
y_{\nu}=\left(  \mathrm{Re}\lambda_{t}+i\mathrm{Im}\lambda_{t}\right)
X_{t}+\left|  V_{us}\right|  ^{4}\mathrm{Re}\lambda_{c}P_{u,c}\;,
\label{NUCoeff}%
\end{equation}
with $\lambda_{q}=V_{qs}^{\ast}V_{qd}$, $X_{t}=1.464\pm0.041$\cite{BurasGHN05}%
, $P_{u,c}=0.41\pm0.04$ for $m_{c}\left(  m_{c}\right)  =1.30\pm
0.05$\cite{BurasGHN05,IsidoriMS05}, while $y_{7V}$ and $y_{7A}$ are given
by\cite{BuchallaBL96}%
\begin{equation}
y_{7V}\left(  \mu\approx1\;\text{GeV}\right)  =\left(  0.73\pm0.04\right)
\,\mathrm{Im}\lambda_{t},\;\;y_{7A}\left(  M_{W}\right)  =\left(
-0.68\pm0.03\right)  \,\mathrm{Im}\lambda_{t}\;. \label{y7AV}%
\end{equation}
In the final numerical applications, we will also use the CKM parameters as
obtained in Ref.\cite{CKMfitter} (compatible with Ref.\cite{UTFit}), and
$V_{us}$ from Ref.\cite{Moulson06}:%
\begin{equation}
\operatorname{Im}\lambda_{t}=1.313_{-0.063}^{+0.112},\;\operatorname{Re}%
\lambda_{t}=-3.09_{-0.22}^{+0.13},\;\operatorname{Re}\lambda_{c}%
=-0.22091_{-0.00093}^{+0.00092},\;\left|  V_{us}\right|  =0.2245\pm0.0016
\label{CKM}%
\end{equation}

All the short-distance physics is encoded in the Wilson coefficients. The
remaining task for computing the branching ratios is thus to get the
matrix-elements of these operators, and to carry out the phase-space
integration. For later use, we collect below the parametrizations of the
branching ratios in terms of the form-factors. For that, we adopt the standard
$K_{\ell3}$ parametrization in terms of the form-factor values at the origin
$(T^{2}=0)$ and of their derivatives there, i.e. their slopes. In this
respect, remember also that $f_{-}^{K^{i}\pi^{j}}$ can be eliminated in favor
of the scalar form-factors (matrix-elements of the scalar current $\bar
{q}\lambda_{a}q$):%
\begin{equation}
f_{0}^{K^{i}\pi^{j}}\left(  T^{2}\right)  =f_{+}^{K^{i}\pi^{j}}\left(
T^{2}\right)  +\frac{T^{2}}{M_{K^{i}}^{2}-M_{\pi^{j}}^{2}}f_{-}^{K^{i}\pi^{j}%
}\left(  T^{2}\right)  \;,
\end{equation}
such that only the vector form-factors at the origin are needed, $f_{0}%
^{K^{i}\pi^{j}}\left(  0\right)  =f_{+}^{K^{i}\pi^{j}}\left(  0\right)  $.

\subsubsection*{Decay rates}

The $K_{\ell3}$ decay rates are%
\begin{equation}
\Gamma\left(  K^{i}\rightarrow\pi^{j}\ell^{+}\nu_{\ell}\left(  \gamma\right)
\right)  =C_{ij}^{2}\frac{G_{F}^{2}S_{EW}M_{K^{i}}^{5}}{192\pi^{3}}\left|
V_{us}\times f_{+}^{K^{i}\pi^{j}}\left(  0\right)  \right|  ^{2}%
\mathcal{I}_{\ell}^{ij}\left(  1+2\Delta_{\ell,EM}^{ij}\right)  \;,
\label{KL3BR}%
\end{equation}
with the Clebsch-Gordan coefficients $C_{0+}=1$ and $C_{+0}=1/\sqrt{2}$. The
Fermi constant $G_{F}$ is fixed from $\mu$ decay as $G_{F}=1.166371\left(
6\right)  \cdot10^{-11}\;$MeV$^{-2}$\cite{MuLan}. Correspondingly, the factor
$S_{EW}=1.0232$ stands for the short distance part of the corrections to the
semi-leptonic weak charged current, to leading order in $m_{Z,W}%
\rightarrow\infty$\cite{Sirlin82}. The phase-space integrals $\mathcal{I}%
_{\ell}^{ij}$, which are functions of the form-factor slopes, are given below.

Adopting a similar parametrization, the $K^{+}\rightarrow\pi^{+}\nu\bar{\nu}$
and $K_{L}\rightarrow\pi^{0}\nu\bar{\nu}$ branching ratios are written as%
\begin{align}
\mathcal{B}\left(  K^{+}\rightarrow\pi^{+}\nu\bar{\nu}\left(  \gamma\right)
\right)   &  =\kappa_{\nu}^{+}\left(  1+\Delta_{EM}\right)  \left|
\frac{y_{\nu}}{\left|  V_{us}\right|  ^{5}}\right|  ^{2},\;\mathcal{B}\left(
K_{L}\rightarrow\pi^{0}\nu\bar{\nu}\right)  =\kappa_{\nu}^{L}\left(
\frac{\operatorname{Im}y_{\nu}}{\left|  V_{us}\right|  ^{5}}\right)
^{2}\;,\label{KnunuBR}\\
\kappa_{\nu}^{+,L}  &  =\tau_{+,L}\frac{G_{F}^{2}M_{K^{+,0}}^{5}\alpha\left(
M_{Z}\right)  ^{2}}{256\pi^{5}\sin^{4}\theta_{W}}\left|  V_{us}\right|
^{8}\left|  V_{us}\times f_{+}^{K^{+,0}\pi^{+,0}}\left(  0\right)  \right|
^{2}\mathcal{I}_{\nu}^{+,0}\;. \label{Coeffnu}%
\end{align}
We will further use\cite{BurasGHN05}%
\begin{equation}
\alpha_{\overline{MS}}\left(  M_{Z}\right)  ^{-1}=127.9,\;\left(  \sin
^{2}\theta_{W}\right)  _{\overline{MS}}=0.231\;, \label{EW}%
\end{equation}
when computing $\kappa_{\nu}^{+,L}$. The experimental errors on $\alpha
_{\overline{MS}}\left(  M_{Z}\right)  $ and $\left(  \sin^{2}\theta
_{W}\right)  _{\overline{MS}}$ as well as the uncertainties related to
(unknown) higher-order, purely electroweak short-distance corrections are
understood to be accounted for in the Wilson coefficients\cite{BuchallaB97}.
Finally, the direct CP-violating contribution to the $K_{L}\rightarrow\pi
^{0}\ell^{+}\ell^{-}$ decays, which arises from the $Q_{7A}$ and $Q_{7V}$
operators (for the general structure and a description of the indirect
CP-violating and CP-conserving contributions, see for example \cite{Kpill}),
takes the form%
\begin{gather}
\mathcal{B}\left(  K_{L}\rightarrow\pi^{0}\ell^{+}\ell^{-}\right)
_{\mathrm{DCPV}}=\left|  \frac{\operatorname{Im}y_{7V}}{\left|  V_{us}\right|
^{5}}\right|  ^{2}\kappa_{\ell}^{V}+\left|  \frac{\operatorname{Im}y_{7A}%
}{\left|  V_{us}\right|  ^{5}}\right|  ^{2}\kappa_{\ell}^{A}\;,\\
\kappa_{\ell}^{i}=\tau_{L}\frac{G_{F}^{2}M_{K^{0}}^{5}\alpha\left(
M_{Z}\right)  ^{2}}{384\pi^{3}}\left|  V_{us}\right|  ^{8}\left|  V_{us}\times
f_{+}^{K^{0}\pi^{0}}\left(  0\right)  \right|  ^{2}\mathcal{I}_{\ell}^{i}\;.
\label{Coeffll}%
\end{gather}

In all cases, the long-distance electromagnetic corrections are moved into the
$\Delta_{\ell,EM}^{ij}$'s, which thus include both virtual photon exchanges
and real photon emissions. Because of the latter, these corrections depend on
the experimental requirements enforced on the kinematics of the corresponding
radiative decay.

For $K_{\ell3}$, also the local QED corrections (including the electromagnetic
and semi-leptonic counterterms) and the QED corrections to the phase-space are
understood in $\Delta_{\ell,EM}^{ij}$. These have been computed both
restricting the $K_{\ell3\gamma}$ phase-space to the three-body
kinematics\cite{Cirigliano01,Cirigliano04}, and for the fully inclusive
case\cite{Inclusive,Neufeld07}.

For $K^{+}\rightarrow\pi^{+}\nu\bar{\nu}$, the correction factor $\Delta_{EM}$
obviously concerns the hadronic part only, and will be computed later on.
Finally, for $K_{L}\rightarrow\pi^{0}\ell^{+}\ell^{-}$, long-distance QED
radiative corrections concern only the lepton pair and are neglected.

\subsubsection*{Phase-space integrals}

We use the quadratic and linear parametrizations for the vector and scalar
form-factors, respectively:%
\begin{equation}
f_{+}^{K^{i}\pi^{j}}\left(  q^{2}\right)  =f_{+}^{K^{i}\pi^{j}}\left(
0\right)  \left(  1+\lambda_{+}^{ij\prime}\frac{q^{2}}{m_{\pi^{\pm}}^{2}%
}+\lambda_{+}^{ij\prime\prime}\frac{q^{4}}{2m_{\pi^{\pm}}^{4}}\right)
,\;\;f_{0}^{K^{i}\pi^{j}}\left(  q^{2}\right)  =f_{+}^{K^{i}\pi^{j}}\left(
0\right)  \left(  1+\lambda_{0}^{ij}\frac{q^{2}}{m_{\pi^{\pm}}^{2}}\right)
\;.
\end{equation}
Then, the phase-space integrals $\mathcal{I}^{ij}$ depends only on the slopes
$\lambda^{ij}$. Explicitly, the $K_{\ell3}$ phase-space integrals are $\left(
i,j=+,0\text{ or }0,+\right)  $%
\begin{equation}
\mathcal{I}_{\ell}^{ij}=\int_{r_{\ell}^{2}}^{\left(  1-r_{\pi}\right)  ^{2}%
}dz\left(  1-\frac{r_{\ell}^{2}}{z}\right)  ^{2}\lambda_{\pi}\left(  \left|
\frac{f_{+}^{K^{i}\pi^{j}}\left(  z\right)  }{f_{+}^{K^{i}\pi^{j}}\left(
0\right)  }\right|  ^{2}\lambda_{\pi}^{2}\left(  1+\frac{r_{\ell}^{2}}%
{2z}\right)  +\frac{3}{2}\left|  \frac{f_{0}^{K^{i}\pi^{j}}\left(  z\right)
}{f_{+}^{K^{i}\pi^{j}}\left(  0\right)  }\right|  ^{2}\frac{r_{\ell}^{2}}%
{z}\left(  1-r_{\pi}^{2}\right)  ^{2}\right)  \;, \label{PSKl3}%
\end{equation}
with $z\equiv q^{2}/M_{K^{i}}^{2}$, $r_{\ell,\pi}\equiv m_{\ell,\pi^{j}%
}/M_{K^{i}}$, $\lambda_{\pi}\equiv\lambda^{1/2}\left(  1,z,r_{\pi}^{2}\right)
$, while those for the rare $K$ decays are
\begin{subequations}
\label{PSrareK}%
\begin{align}
\mathcal{I}_{\nu}^{i}  &  =\int_{0}^{\left(  1-r_{\pi}\right)  ^{2}}%
dz\lambda_{\pi}^{3}\left|  \frac{f_{+}^{K^{i}\pi^{i}}\left(  z\right)  }%
{f_{+}^{K^{i}\pi^{i}}\left(  0\right)  }\right|  ^{2}\;\;(i=+,L),\\
\mathcal{I}_{\ell}^{V}  &  =\int_{4r_{\ell}^{2}}^{\left(  1-r_{\pi}\right)
^{2}}\lambda_{\pi}^{3}\beta_{\ell}\left(  1+\frac{2r_{\ell}^{2}}{z}\right)
\left|  \frac{f_{+}^{K^{0}\pi^{0}}\left(  z\right)  }{f_{+}^{K^{0}\pi^{0}%
}\left(  0\right)  }\right|  ^{2},\\
\mathcal{I}_{\ell}^{A}  &  =\int_{4r_{\ell}^{2}}^{\left(  1-r_{\pi}\right)
^{2}}\lambda_{\pi}\beta_{\ell}\left(  \lambda_{\pi}^{2}\left(
1-\frac{4r_{\ell}^{2}}{z}\right)  \left|  \frac{f_{+}^{K^{0}\pi^{0}}\left(
z\right)  }{f_{+}^{K^{0}\pi^{0}}\left(  0\right)  }\right|  ^{2}%
+\frac{6r_{\ell}^{2}}{z}\left(  1-r_{\pi}^{2}\right)  ^{2}\left|
\frac{f_{0}^{K^{0}\pi^{0}}\left(  z\right)  }{f_{+}^{K^{0}\pi^{0}}\left(
0\right)  }\right|  ^{2}\right)  \;.
\end{align}
with $\beta_{\ell}\equiv\sqrt{1-4r_{\ell}^{2}/z}$ (see Appendix A for
numerical expressions).

\section{Vector form-factors for $K\rightarrow\pi$ transitions}

As we will see in the next section, devoted to the numerical analysis, the
accuracy of $K_{\ell3}$ data are now below the percent level. To make full use
of these impressive results, the theoretical analysis has to reach a
corresponding level of precision. This is not an easy task as it requires some
control over the $\mathcal{O}(p^{6})$ corrections. Fortunately, for the
purpose of estimating FCNC form-factors from those extracted from $K_{\ell3} $
data, two well-chosen ratios are sufficient\cite{MarcianoP96}. Considering
ratios has the immediate advantage that only isospin-breaking corrections,
proportional to $\varepsilon^{(2)}\sim(m_{u}-m_{d})/m_{s}$, have to be brought
under control.

Therefore, in a first stage, we have extended the $\mathcal{O}\left(
p^{4}\varepsilon^{(2)};p^{2}\alpha\right)  $ analysis of
Ref.\cite{Cirigliano01} to the FCNC form-factors. In a second stage, the local
corrections of $\mathcal{O}(p^{6}\varepsilon^{(2)})$ to the ratios will be
discussed. The good surprise is that the structure of the $K\rightarrow\pi$
vector transitions somehow protects our two ratios from large corrections.

\subsection{The vector form-factors at $\mathcal{O}\left(  p^{4}%
\varepsilon^{(2)};p^{2}\alpha\right)  $}

For the $\langle\pi^{i}|V^{j}|K^{k}\rangle$ vector form-factors at
$\mathcal{O}\left(  p^{4}\varepsilon^{(2)};p^{2}\alpha\right)  $, one has to
compute meson and photon one-loop diagrams, with vertices coming from the
leading $\mathcal{O}\left(  p^{2}\varepsilon^{(2)}\right)  $ chiral Lagrangian
and from the $\mathcal{O}(p^{0}\alpha)$ Dashen term $Z_{em}e^{2}%
F^{4}\left\langle U^{\dagger}QUQ\right\rangle $\cite{EckerGPR89}.
Renormalization is carried out by adding the contributions from the
$\mathcal{O}(p^{4}\varepsilon^{(2)})$ strong counterterms $L_{i}%
$\cite{GasserLLi} and from the $\mathcal{O}(p^{2}\alpha)$ QED counterterms
$K_{i}$\cite{VirtualPhotons}. In all these Lagrangians, the CC and FCNC vector
currents are introduced through the covariant derivative with a $\lambda
_{4}\pm i\lambda_{5}$ and $\lambda_{6}\pm i\lambda_{7}$ flavor structure,
respectively, or directly through their contributions to the field-strengths.

In our computation, we keep the vector currents as external sources, to be
coupled later to the lepton pairs, so that the semi-leptonic
counterterms\cite{SemiLeptonic} are not needed to get a UV-finite result. In
that respect, the situation is different for $K_{\ell3}$ and rare $K$ decays.
For the former, these counterterms are ultimately needed to renormalize the
QED corrections where a photon is exchanged between the lepton and the meson.
Altogether, these form the long-distance QED corrections and are included in
$\Delta_{EM}$. On the other hand, for rare $K$ decays, there is no such photon
exchange, and no semi-leptonic counterterm is needed. Further, it should be
clear that short-distance QED corrections are to be accounted for in the
Wilson coefficients of the effective operators. In particular, the $S_{EW}$
factor does not occur for rare $K$ decays (see also the discussion in
Refs.\cite{MarcianoP96,Cirigliano01}).

A straightforward computation gives the following $\langle\pi^{i}|V^{j}%
|K^{k}\rangle$ matrix-elements, including terms up to $\mathcal{O}\left(
p^{4}\varepsilon^{(2)};p^{2}\alpha\right)  $:
\end{subequations}
\begin{subequations}
\label{fV}%
\begin{align}
f_{+}^{K^{0}\pi^{0}}\left(  q^{2}\right)   &  =1+3H_{K^{0}}^{\eta}+H_{K^{0}%
}^{\pi^{0}}+2H_{K^{+}}^{\pi^{+}}-\sqrt{3}\varepsilon^{(2)}(1+H_{K^{0}}^{\eta
}+3H_{K^{0}}^{\pi^{0}}+2H_{K^{+}}^{\pi^{+}})-\sqrt{3}\varepsilon
^{(4)}\;,\label{f00}\\
f_{+}^{K^{+}\pi^{0}}\left(  q^{2}\right)   &  =1+3H_{K^{+}}^{\eta}+H_{K^{+}%
}^{\pi^{0}}+2H_{K^{0}}^{\pi^{+}}+\sqrt{3}\varepsilon^{(2)}(1+H_{K^{+}}^{\eta
}+3H_{K^{+}}^{\pi^{0}}+2H_{K^{0}}^{\pi^{+}})+\sqrt{3}\varepsilon
^{(4)}-\frac{\alpha}{2\pi}\delta_{EM}^{K^{+}}\;,\label{fp0}\\
f_{+}^{K^{0}\pi^{+}}\left(  q^{2}\right)   &  =1+3H_{K^{+}}^{\eta}+H_{K^{+}%
}^{\pi^{0}}+2H_{K^{0}}^{\pi^{+}}+2\sqrt{3}\varepsilon^{(2)}(H_{K^{+}}^{\pi
^{0}}-H_{K^{+}}^{\eta})-\frac{\alpha}{2\pi}\delta_{EM}^{\pi^{+}}%
\;,\label{f0p}\\
f_{+}^{K^{+}\pi^{+}}\left(  q^{2}\right)   &  =1+3H_{K^{0}}^{\eta}+H_{K^{0}%
}^{\pi^{0}}+2H_{K^{+}}^{\pi^{+}}-2\sqrt{3}\varepsilon^{(2)}(H_{K^{0}}^{\pi
^{0}}-H_{K^{0}}^{\eta})-\frac{\alpha}{2\pi}J_{EM}\left(  q^{2},M_{K^{+}}%
^{2},M_{\pi^{+}}^{2}\right)  \;, \label{fpp}%
\end{align}
with
\end{subequations}
\begin{equation}
\delta_{EM}^{i}=1-16\pi^{2}K_{12}^{r}+\log\frac{M_{\gamma}}{\mu}-\frac{3}%
{4}\log\frac{M_{i}}{\mu}\;.
\end{equation}
The loop functions are given in the appendix. The two-point functions
$H_{i}^{j}$ involve $L_{9}$, while the three-point function $J_{EM}$ is free
of counterterms but IR-divergent. The isospin-breaking terms coming from
$\pi^{0}-\eta$ mixing are ($2\hat{m}=m_{d}+m_{u}$)%
\begin{equation}
\varepsilon^{(2)}=\frac{\sqrt{3}}{4}\frac{m_{d}-m_{u}}{m_{s}-\hat{m}%
}=0.01061\pm0.00083\;, \label{e2}%
\end{equation}
and $\varepsilon^{(4)}=\varepsilon_{IB}^{(4)}+\varepsilon_{EM}^{(4)}$
with\cite{GasserL85,EckerMNP99}%
\begin{align}
\varepsilon_{IB}^{(4)}  &  =\varepsilon^{(2)}\frac{\Delta(2A_{K}-A_{\pi
}+A_{\eta}+2M_{K}^{2})-2M_{\pi}^{2}(A_{K}-A_{\pi})-1024\pi^{2}\Delta
^{2}(3L_{7}^{r}+L_{8}^{r})}{24\pi^{2}F_{\pi}^{2}(M_{\eta}^{2}-M_{\pi}^{2}%
)}\nonumber\\
&  =\varepsilon^{(2)}\left(  0.27\pm0.09\right)  =\left(  3\pm1\right)
\cdot10^{-3}\;,\label{e4IB}\\
\varepsilon_{EM}^{(4)}  &  =\alpha\frac{-9Z_{em}(M_{K}^{2}+A_{K})+16\pi
^{2}((2M_{K}^{2}+M_{\pi}^{2})Q_{1}^{r}-6M_{\pi}^{2}Q_{2}^{r})}{18\sqrt{3}%
\pi(M_{\eta}^{2}-M_{\pi}^{2})}=\left(  1\pm2\right)  \cdot10^{-4}\;,
\label{e4EM}%
\end{align}
where $\Delta=M_{K}^{2}-M_{\pi}^{2}$, $A_{i}=M_{i}^{2}\log\left(  M_{i}%
^{2}/\mu^{2}\right)  $, $Q_{1}^{r}=3K_{4}^{r}-6K_{3}^{r}+2K_{5}^{r}+2K_{6}%
^{r}$, $Q_{2}^{r}=K_{9}^{r}+K_{10}^{r}$, and $Z_{em}=(M_{\pi^{\pm}}^{2}%
-M_{\pi^{0}}^{2})/2e^{2}F_{\pi}^{2}\approx0.8$. Numerical estimates are taken
from Ref.\cite{Cirigliano04} (which is based on the quark-mass analysis of
Ref.\cite{Leutwyler96} as well as from general dimensional analysis arguments
for the $K_{i}$ counterterms).

Adopting the parametrization introduced in the first section, the QED
corrections $\delta_{EM}^{i}$ and $J_{EM}$ are moved into the $\Delta_{EM}$
corrections (as well as the small $\varepsilon_{EM}^{(4)}$ effects), and will
thus be dropped from Eqs.(\ref{fV}). It should be clear though that not all
QED effects are moved into $\Delta_{EM}$, since the physical charged particle
masses occur inside the loop functions. In this respect, the separation into
long-distance QED corrections and ``purely'' strong parts is somewhat ambiguous.

From the general expressions, the isospin-breaking corrections for the slopes
and for the form-factors at the origin can easily be obtained. The former will
be discussed later on. For the latter, setting $q^{2}=0$ in Eqs.(\ref{fV}),
one finds
\begin{subequations}
\label{fV0}%
\begin{align}
f_{+}^{K^{0}\pi^{0}}\left(  0\right)   &  =0.9775-\sqrt{3}(0.963\varepsilon
^{(2)}+\varepsilon^{(4)})\;,\label{f000}\\
f_{+}^{K^{+}\pi^{0}}\left(  0\right)   &  =0.9773+\sqrt{3}(0.963\varepsilon
^{(2)}+\varepsilon^{(4)})\;,\label{fp00}\\
f_{+}^{K^{0}\pi^{+}}\left(  0\right)   &  =0.9773-0.0250\varepsilon
^{(2)}\;,\label{f0p0}\\
f_{+}^{K^{+}\pi^{+}}\left(  0\right)   &  =0.9775+0.0257\varepsilon^{(2)}\;,
\label{fpp0}%
\end{align}
where $M_{\eta}$ is set to its physical value in the loop functions (as
prescribed by unitarity). Setting instead
\end{subequations}
\begin{equation}
M_{\eta}^{2}=(2M_{K^{0}}^{2}+2M_{K^{\pm}}^{2}+M_{\pi^{0}}^{2}-2M_{\pi^{\pm}%
}^{2})/3\;, \label{etath}%
\end{equation}
as prescribed at this order, shifts all the form-factors down by a common
$0.0004$.

Inserting $\varepsilon^{(2)},\varepsilon^{(4)}$ as given in Eqs.(\ref{e2}%
,\ref{e4IB}) in Eqs.(\ref{fV0}), very precise estimates can be obtained.
Unfortunately, the theoretical error on the isospin-symmetric $\mathcal{O}%
(p^{6})$ correction is significantly larger than the experimental errors on
the $K_{\ell3}$ data. Specifically, the isospin-symmetric $\mathcal{O}(p^{6})$
result of Ref.\cite{BijnensT03} is, using the numerical estimates for the
local terms found in Ref.\cite{LeutwylerR84} (see also
Refs.\cite{deltaCT,CiriglianoEEKPP05}):%
\begin{gather}
\left.  f_{+}^{K\pi}\left(  0\right)  \right|  _{\mathcal{O}(p^{6})}%
=\delta_{loops}+\delta_{CT}=-0.001\pm0.010\;,\label{fp6}\\
\delta_{loops}=0.0146\pm0.0064,\;\;\delta_{CT}=-\frac{8\Delta^{2}}{F_{\pi}%
^{4}}(C_{12}^{r}+C_{34}^{r})=-0.016\pm0.008\;. \label{DeltaCT}%
\end{gather}
One should also keep in mind that the lattice estimate\cite{Lattice} for
$\left.  f^{K\pi}\left(  0\right)  \right|  _{\mathcal{O}(p^{6})}%
^{lattice}=-0.016$, and thus $\delta_{CT}^{lattice}\approx-0.031$. Further, it
is only with this larger $\mathcal{O}(p^{6})$ correction that $V_{us}$, as
extracted from $K_{\ell3}$ data, satisfies the CKM unitarity. This seems to
indicate that the error on $\delta_{CT}$ is under-estimated. As said, we will
circumvent these difficulties by considering only two ratios of form-factors,
to which we now turn.

\subsection{The ratio $r$}

Once the long-distance QED corrections have been factorized into the
$\Delta_{EM}$'s (see Eqs.(\ref{KL3BR},\ref{KnunuBR})), the form-factors
satisfy%
\begin{equation}
r_{0+}\equiv\frac{f_{+}^{K^{+}\pi^{0}}\left(  q^{2}\right)  }{f_{+}^{K^{0}%
\pi^{+}}\left(  q^{2}\right)  }=\frac{f_{+}^{K^{+}\pi^{+}}\left(
q^{2}\right)  }{f_{+}^{K^{0}\pi^{0}}\left(  q^{2}\right)  }=1+\sqrt
{3}(\varepsilon^{(2)}+\varepsilon^{(4)})\;.\label{FullRatio}%
\end{equation}
This relation is valid up to and including $\mathcal{O}\left(  p^{6}%
,p^{4}\varepsilon^{(2)},p^{2}\alpha\right)  $ terms, since the isospin-exact
$\mathcal{O}(p^{6})$ correction drops out in the ratio. The dominant
corrections then come from $\mathcal{O}(p^{6}\varepsilon^{(2)},p^{4}\alpha)$.
Further, being momentum-independent, it tells us that at this order, the
slopes of the CC form-factors are identical, as well as those for the FCNC
form-factors. Since the same relation holds for scalar form-factors, we can
write%
\begin{equation}
\lambda_{+,0}^{CC}\equiv\lambda_{+,0}^{K^{+}\pi^{0}}=\lambda_{+,0}^{K^{0}%
\pi^{+}},\;\lambda_{+,0}^{FCNC}\equiv\lambda_{+,0}^{K^{+}\pi^{+}}%
=\lambda_{+,0}^{K^{0}\pi^{0}}\;.\label{Slopes}%
\end{equation}
From the two equalities in Eq.(\ref{FullRatio}), one can form the double ratio
$r$ which is exactly $1$ at this order:%
\begin{equation}
r\equiv\frac{f_{+}^{K^{+}\pi^{0}}\left(  q^{2}\right)  }{f_{+}^{K^{0}\pi^{+}%
}\left(  q^{2}\right)  }\frac{f_{+}^{K^{0}\pi^{0}}\left(  q^{2}\right)
}{f_{+}^{K^{+}\pi^{+}}\left(  q^{2}\right)  }=1+\mathcal{O}((\varepsilon
^{(2)})^{2})\;.\label{totRatio}%
\end{equation}
All the momentum-dependences cancel to first order in $\varepsilon^{(2)}$ (and
no further expansion in $1/F_{\pi}^{2}$ is needed). This is a striking
prediction of ChPT. It can be understood from the quark diagrams for each
transition, where the $u$ or $d$ spectator quark plays no role at leading
orders, except for $\pi^{0}-\eta$ mixing and of course long-distance QED
corrections moved into $\Delta_{EM}$. In that respect, notice that all the QED
corrections originating from the meson masses, kept in the form-factors, do
also cancel out completely.

At $\mathcal{O}(p^{6}\varepsilon^{(2)})$, though $\pi^{0}-\eta$ mixing still
cancels in $r$, one could start to feel the effect of the spectator quarks.
Naively, only non-local isospin-breaking contributions at two-loops could
generate corrections (for instance ``sunrise''-type graphs). Still, there is a
good possibility that these effects also cancel out in $r$, at least to a
large extent. A full two-loop computation would be required to check this
conjecture, and is beyond our scope. For now, we will be satisfied by looking
only at the behavior of $\mathcal{O}(p^{6}\varepsilon^{(2)})$ local terms for
the vector form-factor (thus computing only tree-level wave-function and
vertex corrections from the $C_{i}$ Lagrangian of Ref.\cite{BijnensCE99}, in
which we renormalize the $C_{i}$ by $F_{\pi}^{-2}$). We observe that while
there is a local, momentum-dependent correction to Eq.(\ref{FullRatio}),%
\begin{align}
r_{0+}  &  =1+\sqrt{3}(\varepsilon^{(2)}+\varepsilon^{(4)}+\varepsilon
^{(6)})+\frac{16\Delta\varepsilon^{(2)}}{\sqrt{3}F_{\pi}^{4}}\left(
\Delta(4C_{12}-6C_{35})+q^{2}\left(  2C_{12}+C_{65}+C_{90}\right)  \right)
+\mathcal{O}\left(  \alpha\right)  \,,\label{rp6}\\
\varepsilon^{(6)}  &  =\varepsilon^{(2)}\frac{128\Delta^{2}(2M_{K}^{2}+M_{\pi
}^{2})}{F_{\pi}^{4}(M_{\eta}^{2}-M_{\pi}^{2})}\left(  \frac{C_{14}+C_{17}}%
{9}+\frac{C_{18}}{3}-C_{19}-2\frac{C_{20}+C_{31}+C_{32}+2C_{33}}{3}\right)
\;,
\end{align}
all these terms cancel out in the double ratio $r$, apart from a small
$\mathcal{O}\left(  \alpha Z_{em}\right)  $ term%
\begin{equation}
r=1-\frac{64\pi\alpha Z_{em}}{F_{\pi}^{2}}\left(  2\Delta C_{12}-q^{2}%
C_{90}\right)  +\mathcal{O}((\varepsilon^{(2)})^{2}) \label{r6}%
\end{equation}
where $C_{12}\sim$ a few $10^{-6}$\cite{CiriglianoEEKPP05}. This indicates
that indeed, there are strong cancellations at play in the ratio $r$ between
isospin-breaking corrections induced by the spectator quark. Therefore, in our
numerical applications, we will take%
\begin{equation}
r=1.0000\pm0.0002\;, \label{finalr}%
\end{equation}
to account for possible $(\varepsilon^{(2)})^{2}$ effects, $\mathcal{O}%
(p^{4}\alpha)$ corrections or residual $\mathcal{O}(p^{6}\varepsilon^{(2)})$
non-local contributions.

\subsection{The ratio $r_{K}$}

The ratio $r_{K}$ is defined as%
\begin{equation}
r_{K}\equiv\frac{f_{+}^{K^{+}\pi^{+}}\left(  0\right)  }{f_{+}^{K^{0}\pi^{+}%
}\left(  0\right)  }=\left(  1.00027\pm0.00008\right)  +\left(  0.051\pm
0.01\right)  \varepsilon^{(2)}\;,
\end{equation}
and parametrizes the isospin-breaking due to the initial kaon only, i.e. it is
not sensitive to $\pi^{0}-\eta$ mixing. Compared to $r$, it cannot be
expressed in simple analytic form, and is defined only for $q^{2}=0$.

The errors are estimated as twice the shift induced by varying the $\eta$ mass
between its physical and theoretical value, Eq.(\ref{etath}). Still, as
discussed in the previous section, the sensitivity to the spectator quarks is
negligible at $\mathcal{O}(p^{4}\varepsilon^{(2)})$, and the above errors
probably underestimate the full $\mathcal{O}(p^{6}\varepsilon^{(2)})$
corrections. As before, a full analysis goes beyond our scope, and we consider
only local contributions which are%
\begin{equation}
\left(  r_{K}\right)  _{CT-\mathcal{O}(p^{6}\varepsilon^{(2)})}=-\varepsilon
^{(2)}\frac{8}{\sqrt{3}}\delta_{CT}-64\pi\alpha Z_{em}\frac{\Delta}{F_{\pi
}^{2}}C_{12}\;.
\end{equation}
We discard the $O\left(  p^{4}\alpha\right)  $ correction, which is about
$10\%$ of the strong one if $C_{12}\approx C_{34}$. Interestingly, the
combination of counterterms $\delta_{CT}$ is exactly the one occurring in the
isospin limit at $\mathcal{O}(p^{6})$ for the form-factor at the origin,
Eq.(\ref{DeltaCT}), but is now suppressed by an additional $\varepsilon^{(2)}$
factor. Using the numerical estimate $\delta_{CT}=-0.016\pm0.016$, with an
inflated error to account for the discrepancy between lattice (and CKM
unitarity) and model estimates as well as for the neglect of $\mathcal{O}%
(p^{6}\varepsilon^{(2)})$ loop contributions, and with $\varepsilon^{(2)}$
from Eq.(\ref{e2}), we get a very precise estimate for $r_{K}$:%
\begin{equation}
r_{K}=\left(  1.00027\pm0.00008\right)  +\left(  0.12\pm0.07\right)
\varepsilon^{(2)}=1.0015\pm0.0007\;, \label{finalrK}%
\end{equation}
which should thus include the bulk of $\mathcal{O}(p^{6}\varepsilon^{(2)})$ effects.

\section{Numerical analysis}

We start in the next subsection with the estimation of the FCNC form-factor
slopes and rare $K$ decay phase-space integrals. Then, in the following
subsection, we estimate the FCNC form-factors at the origin and discuss the
$K_{\ell3}$ experimental situation. These values are then used in the third
subsection to get the rare $K$ decay matrix-elements, and finally, the last
subsection deals with the long-distance QED corrections for $K^{+}%
\rightarrow\pi^{+}\nu\bar{\nu}$.

\subsection{Slopes and phase-space integrals}

The $K_{\ell3}$ form-factor slopes are determined from Dalitz plot analyses
(corrected for long-distance QED corrections). To an excellent approximation,
the $K^{+}$ and $K_{L}$ slopes are identical (see Eq.(\ref{Slopes})). A
best-fit analysis of ISTRA\cite{ISTRA}, KLOE\cite{KLOE,KLOEprelim},
KTeV\cite{KTeV} and NA48\cite{NA48} data was performed recently in
Ref.\cite{Moulson06}, with the result%
\begin{equation}%
\begin{array}
[c]{l}%
\lambda_{+}^{\prime}=\left(  24.82\pm1.10\right)  \cdot10^{-3}\;,\\
\lambda_{+}^{\prime\prime}=\left(  1.64\pm0.44\right)  \cdot10^{-3}\;,\\
\lambda_{0}=\left(  13.38\pm1.19\right)  \cdot10^{-3}\;,
\end{array}
\;\;\rho=\left(
\begin{array}
[c]{ccc}%
1 & -0.95 & 0.32\\
& 1 & -0.43\\
&  & 1
\end{array}
\right)  \;. \label{ExpSlopes}%
\end{equation}
From these slopes, the $K_{\ell3}$ phase-space integrals are known to about
1/3\%, see Table \ref{TableBR}. It should be noted that about a third of these
errors is due to the poor quality of the fit (i.e., to scale
factors)\cite{Moulson06}, originating in the strong experimental disagreement
between $\lambda_{0}$ measurements. This will hopefully disappear in the near future.%

\begin{table}[t] \centering
$%
\begin{tabular}
[c]{ccc}\hline
& BR$\left(  \%\right)  $ & $\mathcal{I}_{\ell}$\\\hline
$K_{e3}^{L}$ & 40.563$\left(  74\right)  $ & 0.15454$\left(  29\right)  $\\
$K_{\mu3}^{L}$ & 27.047$\left(  71\right)  $ & 0.10209$\left(  31\right)  $\\
$K_{e3}^{+}$ & 5.0758$\left(  290\right)  $ & 0.15889$\left(  30\right)  $\\
$K_{\mu3}^{+}$ & 3.3656$\left(  280\right)  $ & 0.10505$\left(  32\right)  $\\
$K_{e3}^{S}$ & 0.07046$\left(  91\right)  $ & 0.15454$\left(  29\right)
$\\\hline
\end{tabular}
\ \ $%
\caption{Result of the best-fit of ISTRA \cite{ISTRA}, KLOE\cite
{KLOE,KLOEprelim},
KTeV\cite{KTeV} and NA48\cite{NA48} data for the $K_{\ell3}$ branching-ratios,
as performed in \cite{Moulson06}, and phase-space integrals for each mode.}
\label{TableBR}
\end{table}%

Let us relate these slopes to those of the FCNC form-factors. From
Eq.(\ref{Slopes}), we know that they are equal (up to the electromagnetic
effects, already included in $\Delta_{EM}$). From Eqs.(\ref{fV}), one can
immediately get the isospin-breaking correction relating the CC and FCNC
form-factor slopes%
\begin{equation}
\frac{\lambda_{+}^{FCNC}}{\lambda_{+}^{CC}}=0.9986\pm0.0002\;.
\end{equation}
This $\mathcal{O}\left(  p^{4}\varepsilon^{(2)}\right)  $ correction is very
small, actually too small to really account for higher order effects. Indeed,
the linear slopes only arise at $\mathcal{O}\left(  p^{4}\right)  $, where
they are dominated by the $L_{9}$ counterterm, as for the pion form-factor
from which $L_{9}$ is fixed. Since $L_{9}$ is dominated by vector-meson
exchanges, it is well-known that large $SU\left(  3\right)  $ corrections at
$\mathcal{O}(p^{6})$ are then needed to account for $m_{\rho}\neq m_{K^{\ast}%
}$. In our case, we of course do not try to fix $L_{9}$ from the pion vector
form-factor, but simply take the slopes as extracted from $K_{\ell3}$ data. We
thus remain with the $\mathcal{O}(p^{6}\varepsilon^{(2)})$ corrections, which
we estimate from the measured $K^{\ast+}-K^{\ast0}$ mass difference as%
\begin{equation}
\frac{\lambda_{+}^{FCNC}}{\lambda_{+}^{CC}}\approx\frac{m_{K^{\ast+}\left(
892\right)  }^{2}}{m_{K^{\ast0}\left(  892\right)  }^{2}}\approx0.990\;.
\end{equation}
Therefore, to estimate the phase-space integrals relevant for rare $K$ decays,
we will rescale both the linear and quadratic slopes by
\begin{equation}
\frac{\lambda_{+}^{FCNC}}{\lambda_{+}^{CC}}=0.990\pm0.005\;.\label{SlopeV}%
\end{equation}
This theoretical uncertainty will be seen to have a smaller impact on the
phase-space integrals than the experimental uncertainties on the slopes themselves.

For the scalar form-factor slope, the experimental information is less
precise. Further, it is not so clear which resonance should play the dominant
role as the presence of the $\kappa$ pole may blur the picture. We therefore
assign a slightly larger error to account for $\mathcal{O}(p^{6}%
\varepsilon^{(2)})$ effects and set
\begin{equation}
\frac{\lambda_{0}^{FCNC}}{\lambda_{0}^{CC}}=0.99\pm0.01\;. \label{SlopeS}%
\end{equation}
Nevertheless, this slope only matters for $K_{L}\rightarrow\pi^{0}\mu^{+}%
\mu^{-}$, and even there, the impact of this additional uncertainty on the
final result will be very limited.

Using the experimental slopes extracted from $K_{\ell3}$, Eq.(\ref{ExpSlopes}%
), we find
\begin{subequations}
\label{PSInt}%
\begin{align}
\mathcal{I}_{\nu}^{+}  &  =0.15269\pm0.00028\pm0.00007\;,\\
\mathcal{I}_{\nu}^{L}\overset{}{=}\mathcal{I}_{e}^{V,A}  &  =0.16043\pm
0.00030\pm0.00008\;,\\
\mathcal{I}_{\mu}^{V}  &  =0.03766\pm0.00010\pm0.00003\;,\\
\mathcal{I}_{\mu}^{A}  &  =0.08624\pm0.00068\pm0.00009\;,
\end{align}
where the first error comes from the experimental error on the slopes, and the
second from the unknown $\mathcal{O}(p^{6}\varepsilon^{(2)})$ terms. The
correlation between $\mathcal{I}_{\mu}^{V}$ and $\mathcal{I}_{\mu}^{A}$ turns
out to be of about $0.1\%$ and can be safely neglected in evaluating
$\mathcal{B}\left(  K_{L}\rightarrow\pi^{0}\mu^{+}\mu^{-}\right)
_{\mathrm{DCPV}}$. Currently, the above errors are dominated by the
experimental error on $\lambda_{+}^{\prime}$ (except $\mathcal{I}_{\mu}^{A}$
for which $\lambda_{0}$ dominates). As a result, a reduction of about $50\%$
for both the $\lambda_{+}^{\prime}$ and $\lambda_{0}$ experimental errors,
leads to a $\sim50\%$ reduction for all the phase-space integral errors.

\subsection{Form-factors at the origin}

The $K_{\ell3}$ experimental branching ratios are given in Table
\ref{TableBR}, as obtained in the best-fit analysis of Ref.\cite{Moulson06}
(together with the $\pi\pi$, $\pi\pi\pi$ modes and the $K_{S,L}$ and $K^{+}$
lifetimes, $\tau_{L}=51.173\left(  200\right)  $ ns, $\tau_{S}=0.08958\left(
5\right)  $ ns and $\tau^{\pm}=12.3840\left(  193\right)  $ ns). From these
values, and after removing the long-distance QED
corrections\cite{Moulson06,Neufeld07}, one finds
\end{subequations}
\begin{subequations}
\label{VF}%
\begin{align}
\left.
\begin{array}
[c]{ll}%
K_{e3}^{L}: & |V_{us}\times f_{+}^{K^{0}\pi^{+}}\left(  0\right)
|=0.21638\left(  55\right)  \smallskip\\
K_{\mu3}^{L}: & |V_{us}\times f_{+}^{K^{0}\pi^{+}}\left(  0\right)
|=0.21678\left(  69\right)  \smallskip\\
K_{e3}^{S}: & |V_{us}\times f_{+}^{K^{0}\pi^{+}}\left(  0\right)
|=0.21554\left(  142\right)  \smallskip
\end{array}
\right\}  \;|V_{us}\times f_{+}^{K^{0}\pi^{+}}\left(  0\right)
|_{\text{\textrm{exp}}}  &  =0.21645\left(  41\right)  \;,\label{Exp0}\\
\left.
\begin{array}
[c]{ll}%
K_{e3}^{+}: & |V_{us}\times f_{+}^{K^{+}\pi^{0}}\left(  0\right)
|=0.22248\left(  73\right)  \smallskip\\
K_{\mu3}^{+}: & |V_{us}\times f_{+}^{K^{+}\pi^{0}}\left(  0\right)
|=0.22314\left(  106\right)
\end{array}
\;\right\}  \;|V_{us}\times f_{+}^{K^{+}\pi^{0}}\left(  0\right)
|_{\text{\textrm{exp}}}  &  =0.22269\left(  60\right)  \;. \label{ExpP}%
\end{align}

The most precise strategy to get the FCNC form-factors at the origin from
these data is to use the ratios $r$ and $r_{K}$ as%
\end{subequations}
\begin{align}
|V_{us}\times f_{+}^{K^{0}\pi^{0}}\left(  0\right)  |  &  =r\,r_{K}%
\frac{|V_{us}\times f_{+}^{K^{0}\pi^{+}}\left(  0\right)
|_{\text{\textrm{exp}}}^{2}}{|V_{us}\times f_{+}^{K^{+}\pi^{0}}\left(
0\right)  |_{\text{\textrm{exp}}}}=0.2107\pm0.0010\;,\label{FCNC00}\\
|V_{us}\times f_{+}^{K^{+}\pi^{+}}\left(  0\right)  |  &  =r_{K}%
\,|V_{us}\times f_{+}^{K^{0}\pi^{+}}\left(  0\right)  |_{\text{\textrm{exp}}%
}=0.2168\pm0.0004\;. \label{FCNCpp}%
\end{align}
In this way, $\mathcal{O}(p^{6}\varepsilon^{(2)})$ corrections are under good
theoretical control. Further, since the error on $r_{K}\left(  r\right)  $
accounts only for $\sim10\%\left(  3\%\right)  $ and $\sim28\%\left(
0\%\right)  $ of the final error, there is much room for improvements on the
experimental side.

\paragraph{Discussion:}

The strategy of using $r$ instead of $r_{0+}$\cite{MarcianoP96} to get
$|V_{us}\times f_{+}^{K^{0}\pi^{0}}\left(  0\right)  |$ is more precise since,
as discussed previously, higher order effects are under better control for $r$
(see Eq.(\ref{rp6}) and (\ref{r6})). On the other hand, the disadvantage of
using $r$ is that we directly rely on $K_{\ell3}^{+}$ data to estimate
isospin-breaking effects.

In this respect, it should be noted that neutral $K_{\ell3}^{0}$ data are in
much better shape than the charged $K_{\ell3}^{+}$ data. Indeed, all the new
generation Kaon experiments (designed for $K^{0}-\bar{K}^{0}$ mixing) have
published the whole set of information relevant for neutral modes, i.e.
branching ratios, $K_{L}$ lifetime and phase-space measurements. In this way,
$V_{us}\times f_{+}^{K^{0}\pi^{+}}\left(  0\right)  $ and its slopes are known
with very high accuracy. At the moment this situation is not shared by the
charged modes, for which the analysis relies on much older data, and for which
the treatment of radiative corrections is unclear. New preliminary
measurements have been announced by NA48 and ISTRA+ for the ratios
$\mathcal{B}(K_{\ell3}^{+})/\mathcal{B}\left(  K^{+}\rightarrow\pi^{+}\pi
^{0}\right)  $, while KLOE has announced preliminary results for the absolute
branching ratios\cite{KLOEprelim}%
\begin{equation}
\mathcal{B}(K_{e3}^{+})=4.965\left(  52\right)  \%,\;\;\mathcal{B}(K_{\mu
3}^{+})=3.233\left(  39\right)  \%\;.
\end{equation}
All these preliminary values are included in the best-fit analysis of
Ref.\cite{Moulson06} (Table \ref{TableBR}). Unfortunately, the old PDG average
for $\mathcal{B}\left(  K^{+}\rightarrow\pi^{+}\pi^{0}\right)  $ is rather
suspicious (see the discussion in Ref.\cite{Moulson06})), while KLOE data are
not precise enough to really compete yet. As a result, the fit to $K_{\ell
3}^{+}$ data of Ref.\cite{Moulson06} has a rather bad $\chi^{2}$. This
situation should soon improve, as both $K^{+}\rightarrow\pi^{+}\pi^{0}$ and
$K_{\ell3}^{+}$ modes are currently under study.

The above comment can be made more quantitative by looking at the evolution of
the experimental determination of the $r_{0+}$ ratio as new $K_{\ell3}^{+}$
data are announced:%
\begin{equation}
(r_{0+})_{\text{\textrm{exp}}}=\frac{|V_{us}\times f_{+}^{K^{+}\pi^{0}}\left(
0\right)  |_{\text{\textrm{exp}}}}{|V_{us}\times f_{+}^{K^{0}\pi^{+}}\left(
0\right)  |_{\text{\textrm{exp}}}}=\left\{
\begin{array}
[c]{ll}%
1.0328\pm0.0039 & \text{Average }K_{\ell3}^{+}\text{ data, before Kaon
2007\cite{Moulson06},}\\
1.0288\pm0.0034 & \text{Average }K_{\ell3}^{+}\text{ data, after Kaon
2007\cite{Moulson06},}\\
1.0140\pm0.0046\;\; & \text{KLOE preliminary }K_{\ell3}^{+}\text{ data
alone\cite{KLOEprelim}. }%
\end{array}
\right.  \label{r0pExp}%
\end{equation}
Within $2\sigma$, these determinations agree with $r_{0+}$ estimated at
$\mathcal{O}\left(  p^{4}\varepsilon^{(2)}\right)  $ from the $\pi^{0}-\eta$
mixing parameters, Eqs.(\ref{e2}--\ref{e4EM}),%
\begin{equation}
(r_{0+})_{\text{\textrm{th}}}=1+\sqrt{3}(\varepsilon^{(2)}+\varepsilon
^{(4)})=1.0238\pm0.0022\;, \label{r0pTh}%
\end{equation}
where errors are added quadratically. The $\mathcal{O}(p^{6}\varepsilon
^{(2)})$ corrections (see Eq.(\ref{rp6})) are accounted for in the above
error, assuming they scale as $\varepsilon^{(6)}/\varepsilon^{(4)}%
\sim\varepsilon^{(4)}/\varepsilon^{(2)}\sim30\%$\footnote{Comparing the local
$\mathcal{O}(p^{6})$ terms $\delta_{CT}$, Eq.(\ref{DeltaCT}), to those for
$\varepsilon^{(6)}$ as well as for the spectator quark effects, Eq.(\ref{rp6}%
), one can note the larger numerical coefficients for the latter. Also, note
that the spectator quark effects in $r_{0+}$ start only at $\mathcal{O}%
(p^{6}\varepsilon^{(2)})$. For these reasons, the naive counting could have to
be slightly amended.}. Therefore, waiting for the $K_{\ell3}^{+}$ situation to
settle, we will also present final estimates for the rare $K$ decay branching
ratios based on the theoretical $r_{0+}$, Eq.(\ref{r0pTh}), together with
$K_{\ell3}^{0}$ data only, corresponding to%
\begin{equation}
|V_{us}\times f_{+}^{K^{0}\pi^{0}}\left(  0\right)  |_{\mathrm{th}%
}=\frac{r\,r_{K}}{\left(  r_{0+}\right)  _{\text{\textrm{th}}}}|V_{us}\times
f_{+}^{K^{0}\pi^{+}}\left(  0\right)  |_{\text{\textrm{exp}}}=0.2118\pm
0.0004_{\text{\textrm{exp}}}\pm0.0005_{\text{\textrm{th}}}\;.
\end{equation}
Note that in principle, if Eq.(\ref{r0pTh}) is assumed to hold, one should
perform the best-fit average of $K_{\ell3}^{0}$ and $K_{\ell3}^{+}$ data to
get $|V_{us}\times f_{+}^{K^{0}\pi^{+}}\left(  0\right)  |_{\text{\textrm{av.}%
}}=0.21669\left(  36\right)  $, and then use this value to estimate both FCNC
form-factors. As this has only a very small impact numerically, and since we
argued that the recent neutral $K_{\ell3}^{0}$ data are in much better shape
than the older $K^{+}$ data, we prefer not to use this average.

\subsection{Coefficients for the rare $K$ decay branching ratios}

For the $\kappa$ coefficients entering the rare $K$ branching ratios,
Eqs.(\ref{Coeffnu}) and (\ref{Coeffll}), one can in principle avoid
propagating the error due to the $K_{L},K^{+}$ lifetimes by using, instead of
Eqs.(\ref{VF}), the averages:
\begin{subequations}
\label{tauVF}%
\begin{align}
(\tau_{L}|V_{us}\times f_{+}^{K^{0}\pi^{+}}\left(  0\right)  |^{2}%
)_{\text{\textrm{exp}}}  &  =0.23978\left(  64\right)  \cdot10^{-8}\;,\\
(\tau_{+}|V_{us}\times f_{+}^{K^{+}\pi^{0}}\left(  0\right)  |^{2}%
)_{\text{\textrm{exp}}}  &  =0.6141\left(  32\right)  \cdot10^{-9}\;.
\end{align}
Importantly, we do not use directly the $K_{\ell3}$ branching ratios, as is
usually done following Ref.\cite{MarcianoP96}, because the radiative
correction factors $\Delta_{\ell,EM}^{ij}$ have to be removed first (see
Eq.(\ref{KL3BR})).

From these averages, the $K_{L}$ decay coefficients can be expressed as%
\end{subequations}
\begin{align}
\kappa_{\nu}^{L}  &  =\frac{G_{F}^{2}M_{K}^{5}\alpha\left(  M_{Z}\right)
^{2}}{256\pi^{5}\sin^{4}\theta_{W}}\left|  V_{us}\right|  ^{8}\,\,\tau
_{L}\,\,\left(  r\,r_{K}\frac{|V_{us}\times f_{+}^{K^{0}\pi^{+}}\left(
0\right)  |_{\text{\textrm{exp}}}^{2}}{|V_{us}\times f_{+}^{K^{+}\pi^{0}%
}\left(  0\right)  |_{\text{\textrm{exp}}}^{\,}}\right)  ^{2}\,\,\mathcal{I}%
_{\nu}^{0}\label{KappaL1}\\
&  =\frac{G_{F}^{2}M_{K}^{5}\alpha\left(  M_{Z}\right)  ^{2}}{256\pi^{5}%
\sin^{4}\theta_{W}}\left|  V_{us}\right|  ^{8}\,\,(\tau_{L}|V_{us}\times
f_{+}^{K^{0}\pi^{+}}\left(  0\right)  |^{2})_{\text{\textrm{exp}}}\,\,\left(
\frac{r\,r_{K}}{r_{0+}}\right)  ^{2}\,\,\mathcal{I}_{\nu}^{0}\;,
\label{KappaL2}%
\end{align}
and similarly for $\kappa_{e,\mu}^{V,A}$. The second form gives the smallest
errors, i.e. optimizes the use of the experimental information. On the other
hand, doing the same for the $K^{+}$ decay would increase the error because
$r_{0+}$ has a larger impact than $\tau_{+}$:%
\begin{align}
\kappa_{\nu}^{+}  &  =\frac{G_{F}^{2}M_{K}^{5}\alpha\left(  M_{Z}\right)
^{2}}{256\pi^{5}\sin^{4}\theta_{W}}\left|  V_{us}\right|  ^{8}\,\,\tau
_{+}\,\,(r_{K}|V_{us}\times f_{+}^{K^{0}\pi^{+}}\left(  0\right)
|_{\text{\textrm{exp}}})^{2}\,\,\mathcal{I}_{\nu}^{+}\label{KappaP1}\\
&  =\frac{G_{F}^{2}M_{K}^{5}\alpha\left(  M_{Z}\right)  ^{2}}{256\pi^{5}%
\sin^{4}\theta_{W}}\left|  V_{us}\right|  ^{8}\,\,(\tau_{+}|V_{us}\times
f_{+}^{K^{+}\pi^{0}}\left(  0\right)  |^{2})_{\text{\textrm{exp}}}\,\,\left(
\frac{r_{K}}{r_{0+}}\right)  ^{2}\,\,\mathcal{I}_{\nu}^{+}\;, \label{KappaP2}%
\end{align}
Numerically, the first form leads to smaller errors.%

\begin{table}[t] \centering
$%
\begin{tabular}
[c]{c|c|c|ccccc|c}\hline
& $\left(  r_{0+}\right)  _{\mathrm{th}}$ & $\left(  r_{0+}\right)
_{\text{\textrm{exp}}}$ & $\tau_{+}$ & $f(0)_{K_{\ell3}}$ & $\mathcal{I}$ &
$r_{K}$ & $r$ & Future\thinspace?\\\hline
$\kappa_{\nu}^{+}$ & $0.5173\pm0.0025$ & $0.5173\pm0.0025$ & 19 & 43 & 21 &
17 & -- & $\pm0.0023$\\
$\kappa_{\nu}^{L}$ & $2.231\pm0.013$ & $2.209\pm0.017$ & -- & 76 & 12 & 10 &
2 & $\pm0.012$\\
$\kappa_{e}^{V}$ & $0.7832\pm0.0044$ & $0.7755\pm0.0058$ & -- & 76 & 12 & 10 &
2 & $\pm0.0044$\\
$\kappa_{\mu}^{V}$ & $0.1838\pm0.0011$ & $0.1821\pm0.0014$ & -- & 71 & 18 &
9 & 2 & $\pm0.0010$\\
$\kappa_{\mu}^{A}$ & $0.4210\pm0.0040$ & $0.4169\pm0.0045$ & -- & 54 & 37 &
7 & 2 & $\pm0.0029$\\\hline
\end{tabular}
\ \ $%
\caption{Final results for the rare $K$ decay rate coefficients, in units of 
$10^{-10}%
(|V_{us}|/0.225)^8$, for the $r_{0+}$ estimates of Eq.(\ref{r0pTh}%
)
 and (\ref{r0pExp}%
) (current average), respectively. For the latter, the approximate 
breakdown of the errors (in \%) is also indicated. The last column shows the 
possible improvements achievable by a 50\% reduction in the $f^{K^+\pi^0}%
_{+}(0)$, 
$\lambda'_+$ and $\lambda_0$ experimental errors.}
\label{TableResult}
\end{table}%

Using in addition the values quoted in Eq.(\ref{EW}) for $\alpha\left(
M_{Z}\right)  $ and $\sin^{2}\theta_{W}$, we get the results given in Table
\ref{TableResult}, for the experimental or theoretical $r_{0+}$ estimates
discussed in the previous Section\footnote{It should be noted that if the
experimental uncertainty on $\sin^{2}\theta_{W}$ was to be included in the
$\kappa_{\nu}^{+}$ and $\kappa_{\nu}^{L}$ coefficients, their errors would
increase by $14\%$ and $11\%$, respectively. Since the present work concerns
isospin-breaking effects, these uncertainties are left aside. Further, as said
in the first section, they should be dealt with, together with higher-order
electroweak effects, at the level of the short-distance Wilson
coefficients\cite{BuchallaB97}.}. The breakdown of the error (in percent) is
given for the $\left(  r_{0+}\right)  _{\text{\textrm{exp}}}$-based estimates,
and shows that the experimental errors (on the $K^{+}$ lifetime $\tau_{+}$, on
the $K_{\ell3}$ form-factors at the origin $f(0)_{K_{\ell3}}$ and on the
phase-space integrals $\mathcal{I}$) completely dominate, so that there is
much room for improvements. In particular, the last column shows the errors
one would get if the errors on $|V_{us}\times f_{+}^{K^{+}\pi^{0}}\left(
0\right)  |_{\text{\textrm{exp}}}$ (from $K_{\ell3}^{+}$, currently the most
limiting), $\lambda_{+}^{\prime}$ and $\lambda_{0}$ were all reduced by a
factor of two.

Finally, of special interest for the search of new physics, the ratio of the
two neutrino modes can be predicted with good accuracy:%
\begin{equation}
\frac{\kappa_{\nu}^{+}}{\kappa_{\nu}^{L}}=\frac{1}{r^{2}}\,\frac{M_{K^{+}}%
^{5}\,\mathcal{I}_{\nu}^{+}\,(\tau_{+}|V_{us}\times f_{+}^{K^{+}\pi^{0}%
}\left(  0\right)  |^{2})_{\text{\textrm{exp}}}}{M_{K^{0}}^{5}\,\mathcal{I}%
_{\nu}^{0}\,(\tau_{L}|V_{us}\times f_{+}^{K^{0}\pi^{+}}\left(  0\right)
|^{2})_{\text{\textrm{exp}}}}=0.2359\pm0.0015\;.
\end{equation}
Further, the error is currently dominated by $(\tau_{+}|V_{us}\times
f_{+}^{K^{+}\pi^{0}}\left(  0\right)  |^{2})_{\text{\textrm{exp}}}$ and could
easily be reduced by a factor of two in the near future.

\subsection{QED radiative corrections}

\begin{figure}
[ptb]
\begin{center}
\includegraphics[
height=1.6284in,
width=2.6593in
]%
{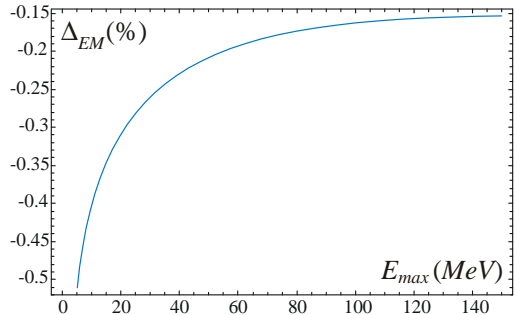}%
\caption{The QED correction to $K^{+}\rightarrow\pi^{+}\nu\bar{\nu}\left(
\gamma\right)  $, in \%, as a function of the maximum energy of the undetected
photon. }%
\label{qedcorr}%
\end{center}
\end{figure}

The final piece needed is the long-distance QED correction to $K^{+}%
\rightarrow\pi^{+}\nu\bar{\nu}$. For that, we have to combine the virtual
photon correction given in Eq.(\ref{fpp}) with the real photon emissions.
These have to be computed at $\mathcal{O}(p^{2}\alpha)$, leading to the
radiative decay rate (see Appendix C for details and explicit expressions):%
\begin{equation}
\Gamma\left(  K^{+}\rightarrow\pi^{+}\nu\bar{\nu}\gamma\right)  =\frac{G_{F}%
^{2}M_{K}^{5}\alpha\left(  M_{Z}\right)  ^{2}\left|  y_{\nu}\right|  ^{2}%
}{256\pi^{5}\sin^{4}\theta_{W}}\frac{\alpha}{\pi}\int_{0}^{\left(  1-r_{\pi
}\right)  ^{2}}dz\lambda_{\pi}^{3}J_{BR}\left(  z,r_{\pi},E_{\max}\right)  \;.
\end{equation}
The function $J_{BR}\left(  z,r_{\pi},E_{\max}\right)  $ accounts for photon
emission with energies up to $E_{\max}$. Altogether, the IR-finite
long-distance QED correction is%
\begin{equation}
\Delta_{EM}\left(  E_{\max}\right)  =\frac{\alpha}{\pi}\frac{1}{\mathcal{I}%
_{\nu}^{+}}\int_{0}^{\left(  1-r_{\pi}\right)  ^{2}}dz\lambda_{\pi}^{3}\left(
J_{BR}\left(  z,r_{\pi},E_{\max}\right)  -J_{EM}\left(  z,r_{\pi}\right)
\right)  \;.
\end{equation}
As shown in Fig.\ref{qedcorr}, $\Delta_{EM}\left(  E_{\max}\right)  $ is of
less than one percent for reasonable $E_{\max}$, while the fully inclusive
correction is $\Delta_{EM}=-0.15\%$, and is thus comparable to those of
$K_{\ell3}$ in magnitude. Still, being free of counterterms at leading order,
it is precisely determined. Looking at Table \ref{TableResult}, it is of the
order of the current uncertainty on $\kappa_{\nu}^{+}$.%

\section{Conclusion}

In this paper, we have presented a theoretical strategy to extract rare $K$
decay matrix-elements from $K_{\ell3}$ data to a few parts per mil. In
summary, from the measured $K_{\ell3}$ branching ratios and form-factor slopes:

\begin{quote}
Step 1: The $K_{\ell3}$ phase-space integrals are directly obtained from the
measured slopes, see Eq.(\ref{PSKl3}) and Appendix A.

Step 2: The $K_{\ell3}$ form-factors at the origin (modulo $V_{us}$, kept as a
parametric uncertainty) are obtained after removing the QED corrections and
the phase-space dependences from the $K_{\ell3}^{L,S,+}$ branching ratios, and
are separately averaged for the charged and neutral $K_{\ell3}$ decays, see
Eqs.(\ref{VF}) and (\ref{tauVF}).
\end{quote}

The outcome of these first two steps is provided by experimentalists, as in
Ref.\cite{Moulson06}, taking into account the correlations among slopes,
branching ratios and QED corrections. Then:

\begin{quote}
Step 3: The slopes are rescaled to account for isospin-breaking,
Eqs.(\ref{SlopeV},\ref{SlopeS}), and rare $K$ decay phase-space integrals are
computed (again, taking care of the correlations among slopes), see
Eqs.(\ref{PSrareK}) and Appendix A.

Step 4: Using the averaged values for the neutral and charged $K_{\ell3}$
form-factors at the origin together with the ratios $r$ and $r_{K}$,
Eqs.(\ref{finalr}) and (\ref{finalrK}), the rare $K$ decay form-factors at the
origin or, even better, directly the coefficients $\kappa$ are obtained
through Eqs.(\ref{KappaL2},\ref{KappaP1}).
\end{quote}

Using this strategy, exceptional control over the hadronic physics is possible
thanks to the two ratios of form-factors $r$ and $r_{K}$, for which
higher-order corrections happen to be very suppressed. Overall, this leads to
a significant improvement over the leading-order analysis of
Ref.\cite{MarcianoP96}. In addition, for the first time, uncertainties due to
higher-order effects in the chiral expansion were carefully studied, allowing
us to carry a thorough error analysis. Finally, QED corrections are accounted
for, both for $K_{\ell3}$ and rare $K$ decays.

The final results for the coefficients entering the rare $K$ decay rates,
Eqs.(\ref{Coeffnu},\ref{Coeffll}), are collected in Table \ref{TableResult}.
As explained in the text, there is at present an open question related to the
isospin-breaking effects observed in the ratio of the $K_{\ell3}^{+}$
form-factor over the $K_{\ell3}^{0}$ one (see Eq.(\ref{r0pExp})). The current
best-fit $K_{\ell3}^{+}$ branching ratios relies on a number of quite old
measurements and has a poor $\chi^{2}$\cite{Moulson06}. This problem will
hopefully disappear in the near future, as the $K^{+}$ modes are currently
under experimental study. Therefore, waiting for the $K_{\ell3}^{+}$ situation
to settle, we consider the values in the first column of Table
\ref{TableResult} as our best estimates.

Compared to the matrix-elements used in Ref.\cite{BurasGHN05}, based on the
leading order analysis of Ref.\cite{MarcianoP96}, the error for $\kappa_{\nu
}^{+}\left(  \kappa_{\nu}^{L}\right)  $ is reduced by a factor of about
$7\left(  4\right)  $, respectively. Using these new values, together with the
Wilson coefficient Eq.(\ref{NUCoeff}) and the CKM parameters Eq.(\ref{CKM}),
we find (adding errors quadratically)%
\begin{equation}%
\begin{array}
[c]{ll}%
\mathcal{B}\left(  K_{L}\rightarrow\pi^{0}\nu\bar{\nu}\right)  =\left(
2.49\pm0.39\right)  \times10^{-11} & \left(  X_{t}:25\%,\;\kappa_{\nu}%
^{L}:3\%,\;\text{CKM}:72\%\right) \\
\mathcal{B}\left(  K^{+}\rightarrow\pi^{+}\nu\bar{\nu}\left(  \gamma\right)
\right)  =\left(  7.83\pm0.82\right)  \times10^{-11}\;\; & \left(
X_{t}:20\%,\;P_{u,c}:31\%,\;\kappa_{\nu}^{+}:3\%,\;\text{CKM}:46\%\right)
\end{array}
\end{equation}
where we have taken $\Delta_{EM}\left(  E_{\max}^{\gamma}\approx
20\,\text{MeV}\right)  \approx-0.30\%$. As the breakdown of the error shows,
the long-distance uncertainties due to the matrix-elements are now negligible
compared to the other (non-parametric) theoretical uncertainties. Similarly,
for the direct CP-violating contribution to $K_{L}\rightarrow\pi^{0}\ell
^{+}\ell^{-}$, the (non-parametric) error is now completely dominated by
short-distance uncertainties in the Wilson coefficients $y_{7A}$ and $y_{7V}$,
Eq.(\ref{y7AV}). However, for these modes, it is the uncertainty over the
indirect CP-violating contribution which is at present the most limiting, and
for that, better measurements of $\mathcal{B}\left(  K_{S}\rightarrow\pi
^{0}\ell^{+}\ell^{-}\right)  $ would be necessary. The impact of our new
values on the $\mathcal{B}\left(  K_{L}\rightarrow\pi^{0}\ell^{+}\ell
^{-}\right)  $ predictions is therefore limited and we refer to \cite{Kpill}
for the numerics.

In conclusion, the hadronic uncertainties due to the matrix-elements of
dimension-six FCNC\ operators are now well below the short-distance
uncertainties in $K\rightarrow\pi\nu\bar{\nu}$ and $\mathcal{B}\left(
K_{L}\rightarrow\pi^{0}\ell^{+}\ell^{-}\right)  _{\mathrm{DCPV}}$. In
addition, being dominated by $K_{\ell3}$ experimental errors, they should
become even more accurate in the near future since it is reasonable to expect
slope measurements and $K_{\ell3}^{+}$ data to further improve. Hadronic
uncertainties will thus not hamper our ability to test the Standard Model with
very high precision.

\subsection*{Acknowledgements}

We would like to thank Gilberto Colangelo, Gino Isidori and St\'{e}phanie
Trine for their comments. Also, special thanks to Uli Haisch for his careful
reading and commenting on the manuscript. This work is partially supported by
the EU contract No. MRTN-CT-2006-035482 (FLAVIAnet). The work of C.S. is also
supported by the Schweizerischer Nationalfonds.\pagebreak 

\appendix               

\section{Phase-space integral coefficients}

Expanding the phase-space integrals in powers of the slopes, and carrying out
the integration, they can be expressed as numerical polynomials%
\begin{equation}
\mathcal{I}=c_{0}+c_{1}\lambda_{+}^{\prime}+c_{2}\left(  \lambda_{+}^{\prime
2}+\lambda_{+}^{\prime\prime}\right)  +c_{3}\lambda_{+}^{\prime\prime2}%
+c_{4}\lambda_{+}^{\prime}\lambda_{+}^{\prime\prime}+c_{5}\lambda_{0}%
+c_{6}\lambda_{0}^{2}\;,
\end{equation}
with the coefficients for each of the nine phase-space integrals collected in
the table below:%
\[%
\begin{tabular}
[c]{c|c|cccc|cc}\hline
& $c_{0}$ & $c_{1}$ & $c_{2}$ & $c_{3}$ & $c_{4}$ & $c_{5}$ & $c_{6}$\\\hline
$\mathcal{I}_{e}^{0+}$ & 0.14084 & 0.4867 & 0.672 & 2.276 & 2.318 & -- & --\\
$\mathcal{I}_{\mu}^{0+}$ & 0.09085 & 0.2938 & 0.474 & 1.784 & 1.751 & 0.2055 &
0.360\\
$\mathcal{I}_{e}^{+0}$ & 0.14480 & 0.5006 & 0.693 & 2.354 & 2.392 & -- & --\\
$\mathcal{I}_{\mu}^{+0}$ & 0.09346 & 0.3023 & 0.489 & 1.846 & 1.809 & 0.2120 &
0.373\\\hline
$\mathcal{I}_{\nu}^{+}$ & 0.13958 & 0.4721 & 0.638 & 2.068 & 2.153 & -- & --\\
$\mathcal{I}_{\nu}^{0},\mathcal{I}_{e}^{V,A}$ & 0.14603 & 0.5156 & 0.729 &
2.586 & 2.572 & -- & --\\
$\mathcal{I}_{\mu}^{V}$ & 0.03092 & 0.2299 & 0.453 & 2.084 & 1.892 & -- & --\\
$\mathcal{I}_{\mu}^{A}$ & 0.07665 & 0.0691 & 0.150 & 0.804 & 0.681 & 0.5526 &
1.198\\\hline
\end{tabular}
\
\]

\section{Loop functions}

Explicit representations for the loop functions in the momentum range
$0<q^{2}<(M_{K^{i}}-M_{\pi^{j}})^{2}$ can be written in terms of
$x=q^{2}/M_{K^{i}}^{2},r_{\pi}=M_{\pi^{j}}/M_{K^{i}}$, $r_{\gamma}=M_{\gamma
}/M_{K^{i}}$ and%
\begin{equation}
z_{1,2}=1-r_{\pi}^{2}\mp x,\;\beta=1+r_{\pi}^{2}-x,\;\;\lambda_{\pi}%
=\sqrt{r_{\pi}^{4}-2\left(  1+x\right)  r_{\pi}^{2}+\left(  1-x\right)  ^{2}%
}\;,
\end{equation}
as
\begin{align}
H_{i}^{j}\left(  x,r_{\pi}\right)   &  =\frac{M_{i}^{2}}{24\left(  4\pi
F_{\pi}\right)  ^{2}}\left(  \frac{z_{2}\lambda_{\pi}^{2}+2(r_{\pi}%
^{2}-\left(  1-x\right)  ^{2})x}{x^{2}}\log r_{\pi}-\frac{\lambda_{\pi}^{3}%
}{2x^{2}}\log\frac{\beta-\lambda_{\pi}}{\beta+\lambda_{\pi}}+\frac{\left(
1-r_{\pi}^{2}\right)  ^{2}}{x}\right. \nonumber\\
&  \;\;\;\;\;\;\;\;\;\;\;\;\;\;\;\;\;\;\;\;\;\;\;\left.  -4\left(  1+r_{\pi
}^{2}\right)  +x\left(  128\pi^{2}L_{9}^{r}-\log\frac{M_{K^{i}}^{2}}{\mu^{2}%
}+\frac{5}{3}\right)  \right)  \;,
\end{align}
and (with $x=q^{2}/M_{K^{\pm}}^{2}$ and $r_{\pi}=M_{\pi^{\pm}}/M_{K^{\pm}}$)%
\begin{align}
J_{EM}\left(  x,r_{\pi}\right)   &  =\left(  2+2\log r_{\gamma}-\log r_{\pi
}\right)  \left(  1+\frac{\beta}{2\lambda_{\pi}}\log\left(  \frac{\beta
-\lambda_{\pi}}{\beta+\lambda_{\pi}}\right)  \right)  +\frac{\beta}%
{4\lambda_{\pi}}C\left(  x,r_{\pi}\right)  \;,\\
C\left(  x,r_{\pi}\right)   &  =\log^{2}\left(  \frac{z_{1}+\lambda_{\pi}}%
{2x}\right)  -\log^{2}\left(  \frac{z_{1}-\lambda_{\pi}}{2x}\right)  +\log
^{2}\left(  \frac{z_{2}-\lambda_{\pi}}{2x}\right)  -\log^{2}\left(
\frac{z_{2}+\lambda_{\pi}}{2x}\right) \nonumber\\
&  +2\log\left(  \frac{z_{1}-\lambda_{\pi}}{z_{1}+\lambda_{\pi}}\right)
\log\left(  \frac{z_{1}+\lambda_{\pi}}{2\lambda_{\pi}}\right)  +2\log\left(
\frac{2\lambda_{\pi}}{z_{2}-\lambda_{\pi}}\right)  \log\left(  \frac{z_{2}%
+\lambda_{\pi}}{2\lambda_{\pi}}\right) \nonumber\\
&  -\log\frac{\lambda_{\pi}^{2}}{r_{\pi}^{2}}\log\left(  \frac{\beta
-\lambda_{\pi}}{\beta+\lambda_{\pi}}\right)  -4\operatorname{Li}_{2}\left(
\frac{\lambda_{\pi}-z_{2}}{2\lambda_{\pi}}\right)  -4\operatorname{Li}%
_{2}\left(  \frac{z_{1}-\lambda_{\pi}}{z_{1}+\lambda_{\pi}}\right)  \;.
\end{align}
The IR-finite remainder obeys $C\left(  0,r_{\pi}\right)  =0$. Also, one can
check that $J_{EM}\left(  0,1\right)  =0$, as required by vector-current conservation.

\section{Bremsstrahlung and QED correction}

For the bremsstrahlung process $K^{+}\left(  P\right)  \rightarrow\pi
^{+}\left(  K\right)  \nu\left(  p_{1}\right)  \bar{\nu}\left(  p_{2}\right)
\gamma\left(  k\right)  $, the $\mathcal{O}\left(  p^{2}e\right)  $ amplitude
is given by%
\begin{equation}
\mathcal{M}^{\mu}=\frac{G_{F}}{\sqrt{2}}\frac{2\,e\,\alpha\left(
M_{Z}\right)  \,y_{\nu}}{2\pi\sin^{2}\theta_{W}}\left(  g^{\mu\alpha
}-\frac{2K^{\alpha}P^{\mu}}{M_{\gamma}^{2}+2K\cdot k}-\frac{2P^{\alpha}K^{\mu
}}{M_{\gamma}^{2}-2P\cdot k}\right)  \bar{u}(p_{\nu})\gamma_{\mu}\left(
1-\gamma^{5}\right)  v\left(  p_{\bar{\nu}}\right)  \;,
\end{equation}
since the vector form-factor is $f_{+}=1$ at this order. Keeping $M_{\gamma
}>0$, this amplitude has to be squared and integrated over the four-body
phase-space, decomposed as%
\begin{equation}
d\Phi_{4}\left(  P;K,k,p_{\nu},p_{\bar{\nu}}\right)  =\frac{dT^{2}}{2\pi}%
d\Phi_{3}\left(  P;K,T,k\right)  d\Phi_{2}\left(  T;p_{\nu},p_{\bar{\nu}%
}\right)  \;.
\end{equation}
The two-body leptonic integral can be done immediately, while the two
variables for the three-body integral are chosen as the photon and pion
energies. Introducing the reduced variables $x=T^{2}/M_{K}^{2}$, $z=2E_{\pi
}/M_{K}$ and $\omega=2E_{\gamma}/M_{K}$, the differential rate is%
\begin{align}
\frac{d\Gamma\left(  x,\omega,z\right)  }{dxd\omega dz}  &  =\frac{G_{F}%
^{2}M_{K}^{5}\alpha\left(  M_{Z}\right)  ^{2}\left|  y_{\nu}\right|  ^{2}%
}{256\pi^{5}\sin^{4}\theta_{W}}\frac{\alpha\frac{{}}{{}}}{\pi\frac{{}}{{}}%
}\left(  \frac{zz_{2}}{\omega}-\frac{\lambda_{\pi}^{2}}{\omega^{2}%
}-\frac{z+\omega}{^{{}}\omega^{{}}}\frac{\left(  z_{1}-\omega\right)  \left(
2-z-\omega\right)  -2xz}{\beta-z-\omega-r_{\gamma}^{2}}\right. \nonumber\\
&  \;\;\;\;\;\;\;\;\;\;\;\;\;\;\;\;\;\;\;\;\;\;\;\;\;\;\;\;\;\;\;\;\;\;\left.
-r_{\pi}^{2}\frac{\left(  2-z-\omega\right)  ^{2}-4x}{\left(  \beta
-z-\omega-r_{\gamma}^{2}\right)  ^{2}}+2z_{2}-z-\omega+r_{\pi}^{2}\right)  \;,
\end{align}
where terms leading to $\mathcal{O}(r_{\gamma})$ contributions to the rate
have been neglected. The integration bounds are%
\begin{gather}
0\leq x\leq\left(  1-r_{\pi}\right)  ^{2},\;\;2r_{\gamma}\leq\omega
\leq1+r_{\gamma}^{2}-\left(  r_{\pi}+\sqrt{x}\right)  ^{2},\;\;a\left(
x,y\right)  -b\left(  x,y\right)  \leq z\leq a\left(  x,y\right)  +b\left(
x,y\right)  \;,\\
a\left(  x,y\right)  =\frac{\left(  2-\omega\right)  \left(  \beta+r_{\gamma
}^{2}-\omega\right)  }{2\left(  1+r_{\gamma}^{2}-\omega\right)  },b\left(
x,y\right)  =\frac{\sqrt{\omega^{2}-4r_{\gamma}^{2}}\sqrt{\left(  z_{1}%
-\omega+r_{\gamma}^{2}\right)  ^{2}-4r_{\pi}^{2}x}}{2\left(  1+r_{\gamma}%
^{2}-\omega\right)  }\;.\nonumber
\end{gather}
There is no need to cut the $x$ integral since the differential rate with
respect to $x$ is, in principle, an observable and is therefore IR-safe.

To account for the experimental constraints, the photon integral will be
limited to $E_{\max}$, i.e. $\omega_{\max}=2E_{\max}/M_{K}$. Since there will
always be some $x$ for which $1+r_{\gamma}^{2}-\left(  r_{\pi}+\sqrt
{x}\right)  ^{2}<\omega_{\max}$, the phase-space boundaries become quite
involved. To simplify them, we make use of the fact that the probability
amplitude becomes imaginary outside the physical phase-space and write%
\begin{equation}
\int_{0}^{\left(  1-r_{\pi}\right)  ^{2}}dx\int_{2r_{\gamma}}^{1+r_{\gamma
}^{2}-\left(  r_{\pi}+\sqrt{x}\right)  ^{2}}d\omega\,\theta\left(
\omega_{\max}-\omega\right)  =\operatorname{Re}\int_{0}^{\left(  1-r_{\pi
}\right)  ^{2}}dx\int_{2r_{\gamma}}^{\omega_{\max}}d\omega\;,
\end{equation}
as can be explicitly checked numerically. When working to leading order in
$\omega_{\max}$, the spurious imaginary part cancels out and after integration
over $z$ and $\omega$, we find the differential rate:%
\begin{equation}
\Gamma\left(  K^{+}\rightarrow\pi^{+}\nu\bar{\nu}\gamma\right)  =\frac{G_{F}%
^{2}M_{K}^{5}\alpha\left(  M_{Z}\right)  ^{2}\left|  y_{\nu}\right|  ^{2}%
}{256\pi^{5}\sin^{4}\theta_{W}}\frac{\alpha\frac{{}}{{}}}{\pi\frac{{}}{{}}%
}\int_{0}^{\left(  1-r_{\pi}\right)  ^{2}}dx\lambda_{\pi}^{3}J_{BR}\left(
x,r_{\pi},E_{\max}\right)  \;,
\end{equation}
where%
\begin{align}
J_{BR}\left(  x,r_{\pi},E_{\max}\right)   &  =-\left(  2+\frac{\beta}{\lambda
}\log\frac{\beta-\lambda_{\pi}}{\beta+\lambda_{\pi}}\right)  \left(
\log\frac{\omega_{\max}}{r_{\gamma}}-\frac{1}{4}\log\left(  \frac{\beta
-\lambda_{\pi}}{\beta+\lambda_{\pi}}\right)  -\frac{1}{2}-\frac{\omega_{\max}%
}{\beta}\right) \nonumber\\
&  -\left(  \frac{1}{2}+\frac{\beta}{\lambda_{\pi}}\right)  \log\left(
\frac{\beta-\lambda_{\pi}}{\beta+\lambda_{\pi}}\right)  +\frac{\beta}{\lambda
}\operatorname{Li}_{2}\left(  \frac{2\lambda_{\pi}}{\lambda_{\pi}-\beta
}\right)  -\frac{2\omega_{\max}}{\beta}+\mathcal{O}\left(  \omega_{\max
}\right)  \;.
\end{align}
As can be checked, combining this with virtual photon corrections, the
IR-divergence indeed cancels out.

For numerical applications, we use the following numerical interpolation for
the full correction (virtual + real photons)%
\begin{gather}
J_{BR}\left(  x,r_{\pi},E_{\max}\right)  -J_{EM}\left(  x,r_{\pi}\right)
=j_{0}\left(  \omega_{\max}\right)  \;\left(  1+x\,j_{1}\left(  \omega_{\max
}\right)  +\mathcal{O}\left(  x^{2}\right)  \right)  \;,\\
j_{0}\left(  a\right)  =0.595+0.965\log a-1.730a+0.201a^{2}+0.114a^{3}%
\;,\nonumber\\
j_{1}\left(  a\right)  =-2.129-0.0711\log a+0.496a+5.977a^{2}-5.585a^{3}%
\;.\nonumber
\end{gather}
It is valid to better than 1\% for all allowed values of $\omega_{\max}$. When
integrated over $x$, it gives an approximation for the total QED contribution
to the inclusive rate to within 5\%, more than sufficient since the QED
correction itself is very small compared to isospin-breaking corrections.


\begin{thebibliography}{9}                                                                                                %

\bibitem {BurasGHN05}A.J. Buras, M. Gorbahn, U. Haisch, U. Nierste, Phys. Rev.
Lett. \textbf{95} (2005) 261805;
A.~J.~Buras, M.~Gorbahn, U.~Haisch and U.~Nierste, JHEP \textbf{0611} (2006)
002.

\bibitem {IsidoriMS05}G. Isidori, F. Mescia, C. Smith, Nucl. Phys.
\textbf{B718} (2005) 319.

\bibitem {Kpill}G. Buchalla, G. D'Ambrosio, G. Isidori, Nucl. Phys.
\textbf{B672} (2003) 387;
G. Isidori, C. Smith, R. Unterdorfer, Eur. Phys. J. \textbf{C36} (2004) 57;
F. Mescia, C. Smith, S. Trine, JHEP \textbf{08} (2006) 088.

\bibitem {MarcianoP96}W.J. Marciano, Z. Parsa, Phys.\ Rev.\ \textbf{D53}
(1996) R1.

\bibitem {BuchallaBL96}G. Buchalla, A.J. Buras, M. E. Lautenbacher, Rev. Mod.
Phys. \textbf{68} (1996) 1125.

\bibitem {CKMfitter}J. Charles \textit{et al.} [CKMfitter Group], Eur. Phys.
J. \textbf{C41} (2005) 1, and Oct. 4, 2006 updated results presented at
ICHEP06 (Moscow) and BEAUTY06 (Oxford).

\bibitem {UTFit}M.~Bona \textit{et al.} [UTfit Collaboration], JHEP
\textbf{0610} (2006) 081.

\bibitem {Moulson06}M.~Moulson [FlaviaNet Working Group on Kaon Decays], talk
given at the 4th International Workshop on the CKM Unitarity Triangle (CKM
2006), Nagoya, Japan, 12-16 Dec 2006, \textit{hep-ex}/0703013;
and update presented by M. Palutan at the Kaon International Conference (KAON
2007), Frascati, Italy, 21-25 May 2007 (see also http://www.lnf.infn.it/wg/vus/).

\bibitem {MuLan}D.~B.~Chitwood \textit{et al.} [MuLan Collaboration],
\textit{hep-ex}/0704.1981.

\bibitem {Sirlin82}A.~Sirlin, Nucl.\ Phys.\ \textbf{B196} (1982) 83.

\bibitem {BuchallaB97}G.~Buchalla and A.~J.~Buras, Phys.\ Rev.\ D \textbf{57}
(1998) 216.

\bibitem {Cirigliano01}V.~Cirigliano, M.~Knecht, H.~Neufeld, H.~Rupertsberger
and P.~Talavera, Eur.\ Phys.\ J.\ \textbf{C23} (2002) 121.

\bibitem {Cirigliano04}V.~Cirigliano, H.~Neufeld and H.~Pichl,
Eur.\ Phys.\ J.\ \textbf{C35} (2004) 53.

\bibitem {Inclusive}V.~Bytev, E.~Kuraev, A.~Baratt and J.~Thompson,
Eur.\ Phys.\ J.\ \textbf{C27} (2003) 57 [Erratum-ibid.\ \textbf{C34} (2004)
523];
T.~C.~Andre, \textit{hep-ph}/0406006.

\bibitem {Neufeld07}H. Neufeld, talk presented at the FlaviaNet Mini-Workshop
on Kaon Decays, Frascati, Italy, 18-19 May 2007 (http://www.lnf.infn.it/wg/vus/).

\bibitem {EckerGPR89}G.~Ecker, J.~Gasser, A.~Pich and E.~de Rafael,
Nucl.\ Phys.\ \textbf{B321} (1989) 311.

\bibitem {GasserLLi}J.~Gasser and H.~Leutwyler, Nucl.\ Phys.\ \textbf{B250}
(1985) 465.

\bibitem {VirtualPhotons}R.~Urech, Nucl.\ Phys.\ \textbf{B433} (1995) 234;
H.~Neufeld and H.~Rupertsberger, Z.\ Phys.\ \textbf{C71} (1996) 131.

\bibitem {SemiLeptonic}M.~Knecht, H.~Neufeld, H.~Rupertsberger and
P.~Talavera, Eur.\ Phys.\ J.\ \textbf{C12} (2000) 469.

\bibitem {GasserL85}J.~Gasser and H.~Leutwyler, Nucl.\ Phys.\ \textbf{B250}
(1985) 517.

\bibitem {EckerMNP99}G.~Ecker, G.~Muller, H.~Neufeld and A.~Pich,
Phys.\ Lett.\ \textbf{B477} (2000) 88.

\bibitem {Leutwyler96}H.~Leutwyler, Phys.\ Lett.\ \textbf{B378} (1996) 313.

\bibitem {BijnensT03}J.~Bijnens and P.~Talavera, Nucl.\ Phys.\ \textbf{B669}
(2003) 341.

\bibitem {LeutwylerR84}H.~Leutwyler and M.~Roos, Z.\ Phys.\ \textbf{C25}
(1984) 91.

\bibitem {deltaCT}V.~Cirigliano, G.~Ecker, H.~Neufeld and A.~Pich, JHEP
\textbf{0306}, 012 (2003);
M.~Jamin, J.~A.~Oller and A.~Pich, JHEP \textbf{0402} (2004) 047.

\bibitem {CiriglianoEEKPP05}V.~Cirigliano, G.~Ecker, M.~Eidemuller, R.~Kaiser,
A.~Pich and J.~Portoles, JHEP \textbf{0504} (2005) 006.

\bibitem {Lattice}D.~Becirevic \textit{et al.}, Nucl.\ Phys.\ \textbf{B705}
(2005) 339;
F.~Mescia, \textit{hep-ph}/0411097;
D.~Becirevic \textit{et al.}, Eur.\ Phys.\ J.\ \textbf{A24S1} (2005) 69;
N.~Tsutsui \textit{et al.} [JLQCD Collaboration], PoS \textbf{LAT2005} (2006)
357;
C.~Dawson, T.~Izubuchi, T.~Kaneko, S.~Sasaki and A.~Soni,
Phys.\ Rev.\ \textbf{D74} (2006) 114502;
D.~J.~Antonio \textit{et al.}, \textit{hep-lat}/0702026.

\bibitem {BijnensCE99}J.~Bijnens, G.~Colangelo and G.~Ecker, JHEP
\textbf{9902} (1999) 020.

\bibitem {ISTRA}O.~P.~Yushchenko \textit{et al.}, Phys.\ Lett.\ \textbf{B581}
(2004) 31;
Phys.\ Lett.\ \textbf{B589} (2004) 111;
V.~I.~Romanovsky \textit{et al.}, \textit{hep-ex}/0704.2052.

\bibitem {KLOE}F.~Ambrosino \textit{et al.} [KLOE Collaboration],
Phys.\ Lett.\ \textbf{B632} (2006) 43;
Phys.\ Lett.\ \textbf{B636} (2006) 166;
Phys.\ Lett.\ \textbf{B636} (2006) 173.

\bibitem {KLOEprelim}Barbara Sciascia [KLOE Collaboration], talk presented at
the Kaon International Conference (KAON 2007), Frascati, Italy, 21-25 May 2007.

\bibitem {KTeV}T.~Alexopoulos \textit{et al.} [KTeV Collaboration],
Phys.\ Rev.\ \textbf{D70} (2004) 092006;
Phys.\ Rev.\ \textbf{D70} (2004) 092007.

\bibitem {NA48}A.~Lai \textit{et al.} [NA48 Collaboration],
Phys.\ Lett.\ \textbf{B602} (2004) 41;
Phys.\ Lett.\ \textbf{B645} (2007) 26;
\textit{hep-ex}/0703002;
J.~R.~Batley \textit{et al.} [NA48/2 Collaboration],
Eur.\ Phys.\ J.\ \textbf{C50} (2007) 329.
\end{thebibliography}
\end{document}